\documentstyle[12pt]{article}

\def\no{\nonumber}
\begin{document}

\begin{titlepage}
\vspace{1cm}

\begin{centering}

{\Large \bf Compensating fields, bosonization and soldering in $QCD_2$ }

\vspace{.5cm}
\vspace{1cm}

{\large Ricardo Amorim and Nelson R. F. Braga}

\vspace{0.5cm}

 Instituto de F\'{\i}sica, Universidade Federal
do Rio de Janeiro,\\
Caixa Postal 68528, 21945-970  Rio de Janeiro, Brazil\\[1.5ex]
\vspace{1cm}

\begin{abstract}
An interesting tool for investigating the quantum features of a field theory is the introduction of compensating fields. For instance, the anomalous divergence of the chiral current can be calculated in the field-antifield formalism from an extended form of $QCD$ with compensating fields.
The interpretation of this procedure from the bosonized point of view, in the two dimensional case, crucially depends on the possibility of defining a bosonized version for the extended theory. We show, by using some recent results on the soldering of bosonized actions corresponding to chiral fermions, how is the mapping between bosonic and fermionic representations of this extended $QCD_2$. 
In the bosonic formulation the anomalous divergence of the chiral current shows up from the 
equations of motion of the compensating fields.

\end{abstract}

\end{centering}

\vspace{1cm}

\noindent PACS: 03.70.+k, 11.10.Ef, 11.15.-q

\vfill

\noindent{\tt amorim@if.ufrj.br,
braga@if.ufrj.br}
\end{titlepage}

\pagebreak

\section{Introduction}
\setcounter{equation}{0}
Compensating fields can be defined as those  that  enlarge the local symmetry content of an action in such a way that the original theory is recovered at  the unitary gauge. This means that classically the number of degrees of freedom is not changed by the introduction of a set of compensating fields as they can  be completely removed by the gauge fixing procedure. This kind of field space enlargement has been used in a large number of works in the context of Hamiltonian\cite{H} as well as Lagrangian\cite{L,AD} descriptions of field theories. A fundamental question arises when  we introduce compensating fields: are the corresponding  new symmetries obstructed by quantum effects? Several examples, within a Lagrangian formalism, show that compensating fields do not add new anomalies \cite{L1,AB,ABH}( in other words, they do not change the BRST cohomology\cite{BRST} ). It can be shown that this is also the case within Hamiltonian descriptions \cite{AT}. In spite of  possible cohomologically trivial contributions to a field theory, compensating fields are an important tool for extracting informations about its quantum features.
Recent examples are those of Refs.\cite{AB,ABH} where a general procedure for calculating anomalous divergences of global currents in the field antifield 
framework was developed. In these articles, the introduction of compensating fields leads to quantum corrections to the master equation and therefore to additional terms  in the quantum action.
These quantum corrections make it possible to extract the anomalous divergences of Noether currents from the sole imposition of independence of the vacuum functional with respect to the gauge fixing.
These results were illustrated with  $QCD$ in four dimensions.

\bigskip

In the present work we would like to explore some consequences of the introduction of compensating fields in $QCD$ in two space-time dimensions\cite{EA}. As it is well known, field theories defined in two dimensions have an  interesting particular aspect: a two dimensional gauge theory with fermionic matter fields  can be mapped into a corresponding theory involving only bosonic  fields.  The structure and the transformation properties of the bosonized version of a fermionic gauge theory depends essentially on the form of the coupling to the gauge field. For example, in  standard $QCD_2$, where the gauge field is coupled to the vector matter current, the corresponding bosonic matter field is gauge invariant. On the other hand, in the chiral $QCD_2$ case, where the gauge field is coupled to the chiral current, the bosonic matter field transforms as an element of the gauge group. Therefore, if one tries to look at the introduction of compensating fields as in  \cite{AB,ABH} from the bosonized point of view, one faces the problem of how to find the appropriate bosonized version of the theory. Considering the case of a pure vector coupling, like $QCD_2$, the corresponding bosonized version is well known. However, the introduction of compensating fields corresponds to coupling an additional pure gauge field to just one of the chiralities. So, the theory including  compensating fields can not be bosonized in the same way.
We will see that this problem can be solved  by generalizing  recent results from ref. \cite{ABW} on how to solder\cite{solder} two chiral fermions, but here with different gauge fields in each chiral sector. 
\bigskip

We will see that when we carefully include in the bosonic formulation the counterterms that solve the quantum master equation at one loop order for the fermionic case, 
we  arrive at a bosonized action
where the compensating fields play the role of collective fields as in \cite{AD}. 
It is interesting to emphasize that it is just these counterterms that
make it possible to have the complete set of symmetries  manifest  in the bosonized version of the theory. 
The introduction  of the counterterms also allows us to derive a complete mapping between 
the chiral currents and their anomalous divergences in the two formulations of the model. It is natural then to interpret some of the equations of motion  in the gauged WZW model
\cite{EA,Wi} as being just the bosonic form of the anomalous divergence of the
Noether chiral currents.
\bigskip

This work is organized as follows: in section (2) we briefly review the ideas of references \cite{AB,ABH} and present the corresponding results for the case of $QCD_2$. In section (3) the Abelian version of the model discussed in
section (2) is bosonized using the soldering techniques. A mapping between the bosonic and fermionic descriptions of the currents and their divergences is also presented.
Section (4) essentially generalizes the results derived in section (3) to the non-Abelian situation. We devote section (5) to some general comments and concluding remarks.

\section{Compensating fields and the gauged chiral symmetry in $QCD_2$}
\setcounter{equation}{0}
In Ref.\cite{ABH} we have shown that any group $G$ of rigid internal transformations of a set of fields $\phi^i$, $i=1,2,..n$, can be used to enlarge the local symmetry content of an action $S_0[\phi^i]$. Local symmetries are introduced when the group elements are promoted to compensating fields in a proper way. To show how this works, let  us consider that the action of $g\in G$ over $\phi^i$ is 

\begin{equation}
\label{1}
\phi^{i'}=\phi^{i'}(\phi,g)
\end{equation}

\noindent where a subsequent transformation of $\phi^{i'}(\phi,g)$ under a group element $h$ can be obtained directly from $\phi^i$ through the action of $hg$. The usual  group axioms are assumed, and  the identity transformation is generated by the unity element {\bf 1} of $G$.
According to (\ref{1}),

\begin{equation}
\label{2}
\phi^{i}=\phi^{i}(\phi^{'},g^{-1})\,.
\end{equation}

Writing $S_0[\phi^i]$ with the aid of (\ref{2}) and dropping the primes enable us to get an action $S_1=S_1[\phi^i,g(x)]$ with a local set of symmetries which comes from a convenient left multiplication of the group elements, now taken as local fields. Applying these ideas to the Yang-Mills action in two dimensions with $G$ as the group of  non-Abelian chiral transformations, $S_1[\phi^i,g(x)]$ can be written as

\begin{equation}
\label{3}
S_1\,[ \psi \, , \, {\overline \psi}\,,\,  A_\mu \, , \, g \, ] \,
=\,\int d^2x \Big( - {1\over 4} Tr ( F^{\mu\nu}\,
F_{\mu\nu} ) + i\overline \psi \gamma^\mu \big(
\partial_\mu -i {\tilde A}_\mu \big) \psi\,\Big)
\end{equation}

\noindent where
\begin{equation}
\label{4}
{\tilde A}_\mu \,= {\tilde A}_\mu [ A_\mu\,,g\,]\,
= \, P_-\,A_\mu \,+\, P_+ \,B_\mu
\end{equation}

In the above expressions, $P_\pm={1\over2}(1\pm\gamma_5)$ are the chiral projectors and the composite field

\begin{equation}
\label{5}
B_\mu\,=\, g^{-1} A_\mu g \,+\, i  g^{-1}\partial_\mu g 
\end{equation}

\noindent corresponds to a finite gauge transformation of the Yang-Mills connection $A_\mu$. We can see that $S_1\,[ \psi \, , \, {\overline \psi}\,,\,  A_\mu \, , \, g \, ]$ is invariant under the set of local gauge
transformations

\begin{eqnarray}
\label{6}
\delta \psi &=& i (\eta(x)\,-\,\epsilon(x) P_+ \,)\psi, \; \; 
\delta {\overline \psi} = - i {\overline \psi} (\eta(x) - \epsilon(x) P_- )\no\\
\delta A_\mu &=& \partial_\mu \eta(x) \,+\, i [\eta(x)\,,\,A_\mu ], \, \,
\delta g = i ( g\epsilon(x) \,+\, [\eta(x)\,,\,g]\,,
\end{eqnarray}

\noindent where $\epsilon\,=\,\epsilon^a T^a\,$, $\eta\,=\,\eta^a T^a$ take values in a
the  $SU(N)\,\,$ algebra, with generators $T^a$ satisfying
$[T^a,T^b]=if^{abc}T^c$, $tr (T^a T^b)=\delta^{ab}$. We assume here that the connections and the fermions belong to the fundamental representation of SU(N). 

Transformations (\ref{6}) close in an algebra: 
$[\delta_1\,,\,\delta_2\,]\,\phi^i\,=\,\delta_3\phi^i$
for any field $\phi^i=\{\psi,\bar\psi,A_\mu,g\}\,$ when the composition rules for the parameters of the transformation  are given by

\begin{eqnarray}
\label{7}
\eta_3 &=& i [\eta_1\,,\,\eta_2\,]\nonumber\\
\epsilon_3 &=& i \Big( [\eta_1\,,\,\epsilon_2]\,
+\,[\epsilon_1\,,\,\eta_2]\,-\,[\epsilon_1\,,\,\epsilon_2]\,\Big)\,\,,
\end{eqnarray}
which shows the semi-direct product character of $SU(N)\times SU(N)$ for the gauge structure found in (\ref{6}).
\bigskip

In the gauge $g={\bf 1}$, $S_1 $ trivially reduces to the usual Yang-Mills action  and the local chiral symmetry is not manifest anymore. Furthermore, for $g={\bf 1}+ i \beta$, we get

\begin{equation}
{\delta S_1\over \delta \beta^a}\vert_{_{\beta=0}}\,= \,
\,\Big( \,D_\mu J^\mu_R\,\Big)^a\,,
\label{8}\end{equation}

\noindent which was the starting point, in ref. \cite{ABH}, for deriving the anomalous divergence of the chiral current
$J^{\mu\,a}_R\,=\,{\overline \psi} \gamma^\mu T^a P_+ \psi\,$.  In (\ref{8}),

\begin{equation}
\label{9}
\Big( D_\mu J^\mu_R\,\Big)^a \,\equiv\,\partial_\mu J^{\mu\,a}_R\,
+\,f^{abc} \,A_\mu^b J^{\mu\,c}_R\,\,.
\end{equation}

 The quantization of such a theory, along the field-antifield formalism \cite{BRST,GPS}, starts by constructing the BV action

\bigskip
\begin{eqnarray}
\label{10}
S &=& S_1 + \int d^2x \Big( i \psi^\ast ( c -b P_+)
\psi-i{\overline \psi}
(c -b P_-){\overline \psi}^\ast
+ Tr\{ i g^\ast ( g\,b
+\, [c\,,\,g]) \nonumber \\ 
&+&  A^\ast_\mu D^\mu c\,+\, {i\over 2} c^\ast [c\,,\,c]\,
-\,
{i\over 2} b^\ast ( [b\,,\, b ]\,-2 [c\,,\,b] )\,\}\,\, \Big)
\end{eqnarray}
where the set of fields $A_\mu, g, \bar\psi$ and $\psi$ has been extended to include ghosts
$c$ and $b$ ( corresponding to 
parameters $\eta$ and $\epsilon$ respectively ) and also 
antifields (sources of BRST variations) associated with each field. 
In (\ref{10}) the brackets represent graded commutators.

As it is well known, the BRST variation of any functional $X(\phi,\phi^*)$ is given by

\begin{equation}
\label{11}
s\,X=(X,S)
\end{equation}
with $S$ given by (\ref{10}) and the antibrackets defined in such a way that for any two local functionals of the fields and antifields  $X$ and $Y$,
 $ \left(X,Y\right) = \frac{ \partial^RX }{ \partial \Phi^A } \frac{ \partial^LY }{ \partial\Phi^*_A } -
 \frac{ \partial^RX }{ \partial \Phi^A_* } \frac{ \partial^LY }{ \partial \Phi_A }$ . In the above equation and when pertinent, we are using the de Witt notation of sum over repeated indices and integration over the corresponding intermediary variables.
By construction, $S$ is BRST invariant, but the functional generator with $S$ as the quantum action will be well defined at one loop order only if $\Delta S=0$ or if $S$ can be extended to some $W=S+\hbar M_1$ such that

\begin{equation}
\label{12}
s\,M_1=i\Delta S\,.
\end{equation}

\noindent In the expressions above
$\Delta ={(-1)}^{ \epsilon_A+1 } \frac{ \partial^R \partial^R }{ \partial\Phi^A \partial\Phi^*_A }$
is a potentially singular operator that must be regularized. If we choose a regularization that keeps the vector symmetry as a preferential one, we get, by using standard procedures, that

\begin{equation}
\label{13}
\Delta\,S=-{{i}\over{4\pi}}Tr\int d^2x\epsilon^{\mu\nu}\left(c\partial_\mu A_\nu-(c-b)\partial_\mu B_\nu\right)
\end{equation}

\noindent where $B_\mu$ is given by (\ref{4}) and $\epsilon^{01}\,=\,-\epsilon^{10}\,=1\,$. We observe that in the $QCD_2$ limit ( $g\rightarrow{\bf 1}$, $b\rightarrow 0$ ) $\Delta S$ vanishes identically. Furthermore, we can show that even without that limit, $g$ and $b$ do not belong to the cohomology at ghost number one and so there exists some $M_1$
that solves (\ref{12}). To see this, we note  from (\ref{10}),(\ref{11})
that

\begin{eqnarray}
\label{14}
s\,g&=&igb+i[c,b]\nonumber\\
sb&=&-ib^2+i[c,b]
\end{eqnarray}

If we represent $g$ as $g=exp( i\Lambda )$, $\Lambda$ taking values in the SU(N) algebra, we can show that

\begin{equation}
\label{15}
s\,\Lambda=b+i[\Lambda,{b\over2}-c]-{1\over12}[\Lambda,[\Lambda,b]]+\dots
\end{equation}
Expressions (\ref{13}) and (\ref{15}) imply that under the linearized BRST
transformations,  $s^1\Lambda=b$, $s^1b=0$, $b$ and $\Lambda$ form a doublet and therefore are absent from the cohomology\cite{BDK,PS}. Actually, one can
verify that

\begin{equation}
\label{16}
M_1=-{i\over{4\pi}}\,tr\,\int_{\partial M} dx^2\,\epsilon^{\mu\nu}\partial_\mu g g^{-1} A_\nu \,\,-\,\, \Gamma [ g ] \,\,,
\end{equation}

\noindent where

\begin{equation}
\label{gamma}
\Gamma [g]\,=\, {1\over 12\pi}\epsilon^{\mu\nu\rho}\,tr\,\int_M d^3 x  \Big( g^{-1}\partial_\mu g
g^{-1}\partial_\nu gg^{-1}\partial_\rho g \,\Big) \,\,,
\end{equation}

\noindent solves (\ref{12}). $\Gamma(g)$ is the Wess-Zumino functional, defined in a 3D manifold $M$ with boundary $\partial M$ representing 2D Minkowiski space-time\cite{Wi}.
\bigskip

To finish this section, we would like to comment that, although the quantum theory is not obstructed by gauge anomalies, it presents an anomalous
divergence of the chiral current, as expected. Constructing a path integral $ Z[J] $ by using $ W=S_1 + \hbar M_1 + $  $ \,gauge \,\, fixing \,\,terms \,\, $ as the quantum action, we can show that the Batalin-Fradkin theorem that implies the independence of $ Z[J] $ with respect to the gauge fixing 
$ g = { \bf 1 } + i \beta $ ( see ( \ref{2} ) ) leads to

\begin{equation} 
\label{17}
< { { \delta W } \over { \delta \beta^a } } >_{_{ \beta^a = 0 } } = \,< ( D_\mu J^\mu_R \, )^a + { 1\over {4\pi} } \epsilon^{ \mu\nu } \partial_\mu A_\nu^a >_{_{ \beta^a = 0 } }\, = \, 0\,\,.
\end{equation}

\noindent This is a non trivial result that comes naturally from the $ QCD_2 $ extension   presented above. Observe that the expected values appearing in (\ref{17}) are calculated within the  $ QCD_2 $ sector.

\section{Bosonization and soldering}
\setcounter{equation}{0}

In this section we will consider some fundamental points related to the bosonization of the Abelian version of the theory
described above. The non-Abelian case will be described in the next section. Defining $ \psi_{R/L}\, =  \,P_\pm \psi $,
we can write the fermionic sector of the abelianized version of (\ref{3}) as

\begin{eqnarray}
\label{31a}
S_F &=& \int d^2 x \left( i\bar\psi_R \gamma^\mu \left(  \partial_\mu - iB_\mu \right) \psi_R
+i \bar\psi_L \gamma^\mu \left( \partial_\mu - iA_\mu \right) \psi_L
\right)  \\
\label{31b}
&\equiv& S^+_F [ \bar\psi_R , B_\mu , \psi_R ]\,\,+\,\,
 S^-_F [  \bar\psi_L , A_\mu , \psi_L ]
\end{eqnarray}

\noindent where $ B_\mu = A_\mu - \partial_\mu \Lambda $ is the Abelian limit of (\ref{5}).
Standard bosonization technics enable us to derive the bosonized versions of each of the chiral actions $\,S^+_F \, $ and $\,S^-_F\,$  respectively 
as\footnote{ $ x^\pm = { 1 \over \sqrt{2} }( x^0 \pm x^1 )\, , \, \partial_\pm \, = \, { 1 \over \sqrt{2} }( \partial_0 \pm \partial_1 ) \, , \,  A^{\pm} \, = \, { 1 \over \sqrt{2} } ( A^0 \pm A^1 ) \, $ }

\begin{equation}
\label{32}
S_+ \,=\,{ 1\over{4\pi } } \int d^2 x \left[ \partial_+ \varphi \partial_- \varphi + 2B_+
\partial_- \varphi + a B_+ B_-  \right]
\end{equation} 

\noindent and

\begin{equation}
\label{33}
S_-\, =\, {1\over{4\pi}} \int d^2 x \left[ \partial_+\rho \partial_-\rho + 2 A_-
\partial_+ \rho + bA_ + A_- \right] \,\,.
\end{equation} 
\bigskip

\noindent In the above expressions, $a$ and $b$ are free parameters representing the arbitrariness in the regularization procedure. 
Observe that each one of these bosonic actions represents the fermionic determinant of the corresponding chirality.  As it is well known, the complete fermionic determinant corresponding to the complete action (\ref{31a}) is not just the product of the two chiral determinants since there are interference terms. As pointed out  in reference\cite{BB}, the complete determinant comes out if we consider the correct Bose symmetrization in the perturbative calculations. 
An equivalent calculation of the complete bosonized action can be done by using soldering technics   \cite{ABW,solder}, which will be described in what follows . If we gauge the following global
symmetries of the action $S_\pm$

\begin{eqnarray}
\label{33a}
\delta \varphi &=& \alpha \,\,\,,\,\,\, \delta B_\pm \,=\, 0
\nonumber\\
\delta \rho &=& \alpha\,\,\,,\,\,\, \delta A_\pm\,= \, 0
\end{eqnarray}

\noindent by using the Noether procedure we can see that the action

\begin{equation}
\label{34}
S\, = \, S_+ \,+\, S_- -\int d^2x \left[ E_+ { \cal J}_- + E_- { \cal J}_+ - { 1\over { 2 \pi }} E_+ E_- \right]
\end{equation}

\noindent with 

\begin{eqnarray}\label{35}
{\cal J}_+&=& {1\over {2\pi}} \left( \partial_+ \varphi + B_+ \right) \nonumber\\
{\cal J}_-&=& {1\over{2\pi}} \left( \partial_- \rho + A_- \right)
\end{eqnarray}

\noindent is now invariant under (\ref{33a}) with $\alpha = \alpha (x) $ a local parameter, if
the transformations of the soldering gauge fields are given by $\delta E_\pm = \partial_\pm \alpha $.

In (\ref{34}) the interference terms depending in $ E_\pm $ are those that solder the $\varphi$ and $\rho$ sectors of the theory. By eliminating $ E_\pm $ with the aid of their equations of motion, we arrive
at an effective form of $S $ given by

\begin{eqnarray}
\label{36}
S^{\prime} &=& {1\over{4\pi}} \int d^2  x \Big[ \partial_+ \,(\,\varphi\,-\,\rho\,) \partial_-\,(\,\varphi\,-\,\rho\,)\, + 2 A_+ \partial_- ( \varphi\,-\,\rho ) \nonumber\\
&-& 2 A_-\partial_+ ( \varphi\,-\,\rho ) \,+\, ( a\,+\,b\,-\,2)\, A_+ A_-\,-\,2 \partial_+\Lambda \partial_- ( \varphi - \rho) \nonumber\\
&+&
\partial_+ \Lambda \partial_- \Lambda \,+\,  \Lambda \Big(  a (\partial_- A_+ \,+\, \partial_+ A_-)\, - 2 \partial_+ A_- \,\Big)
\Big]
\end{eqnarray}

As one can observe, in (\ref{36}) $\varphi $ and $\rho $ combine as the collective field \cite{AD} $\Phi=\varphi-\rho$ and at the same time the linear combination $\Xi=\varphi+\rho$ is absent from the theory.
If we assume that the bosonization has been done keeping the vector symmetry
as a preferential one, we must choose the free parameters as $a=b=1$. This 
condition also comes out if we assume that the invariance under the usual U(1) (Maxwell) 
transformations is not lost. As a consequence of (\ref{6}), U(1) gauge invariance is manifest
when $\delta A_\pm =  \delta B_\pm = \partial_\pm \eta $ and the
other fields remain invariant. This process is the one
that keeps  unitarity and presents the correct bose symmetry, when
a diagramatic analysis of the bosonization procedure is performed \cite{BB}.
We see that (\ref{36}) then reduces to

\begin{equation}
\label{37}
S^{\prime\prime} \,=\, {1\over 4\pi} \int d^2x \Big[ \partial_+ \,( \Phi - \Lambda )
\partial_-\,( \Phi - \Lambda )\,-\, (2 \Phi - \Lambda\,) 
( \partial_- A_+ \, - \,\ partial_+ A_-  )
\Big]
\end{equation}

Performing an analysis of the set of gauge transformations that keep
(\ref{37}) invariant, we see that the soldering symmetry  is trivially 
satisfied due to the definition of $\Phi$. This action is also invariant under the U(1) gauge transformations, but the chiral symmetry ( associated with the parameter $\epsilon$ in the Abelian version of (\ref{6})) is lost, what is not surprising, once the invariances presented by the bosonized action must be related to the quantum symmetries of the corresponding fermionic quantum action.
Actually, if we add to $S^{\prime}$ the counterterm 

\begin{equation}
\label{38}
M_1\,=\, {1\over 4\pi}\,\int d^2 x \Big[ \Lambda ( \partial_- A_+\,-\,\partial_+ A_-  )
 \Big]
\end{equation}

\noindent which is the Abelian limit of (\ref{16}), we see that the total action 

\begin{eqnarray}
\label{39}
W &=&S^{\prime\prime} + \hbar M_1 \nonumber\\
&=& {1\over {4\pi} } \int d^2x \Big[ \partial_+ \,( \Phi - \Lambda )
\partial_- \,( \Phi - \Lambda )\,-\, 2 ( \Phi - \Lambda\,) 
( \partial_- A_+\,-\,\partial_+ A_-  ) \Big]
\nonumber\\
& &
\end{eqnarray} 

\noindent besides soldering and U(1) symmetries, is trivially invariant under
the chiral symmetry

\begin{equation}
\label{40}
\delta \Phi\,=\,\epsilon \,,\,\delta \Lambda\,=\,\epsilon\,,
\end{equation}

\noindent which now survives also  the bosonization and soldering processes.
It is interesting to note that not only $\varphi$ and $\rho$ combine in the
collective field $\Phi$\cite{AD}, but also that $\Lambda$ and $\Phi$ themselves combine
in a second collective field $\bar \Phi = \Phi - \Lambda $. In terms of $ \bar\Phi $,
the bosonized version of the extended  $QED_2$ is identical to the
bosonized version of the standard Schwinger model. As expected, the bosonized version of the theory presents a much more simpler form than its fermionic counterpart, with action given by (\ref{31a}) added to (\ref{38}).
\bigskip

To conclude this section, we would like to consider some aspects associated with the mapping between the currents that appear in both descriptions of the model . 
The matter current is defined as the object that couples to the vector gauge field in the quantum action  

\begin{equation}
\label{Acurr}
J_{\pm}\,\equiv\,{ \delta W \over \delta A_{\mp}}\vert_{_{ \Lambda\,=\,0}}\,\,\,\,,
\end{equation}

\noindent where the condition $\Lambda\,=\,0\,$ fixes the unitary gauge  in order to return to the original description of theory (QED2 in this case).
Thus the fermionic currents are mapped in

\begin{eqnarray}
\label{41}
J_+^F&=& {1\over \sqrt{2}}\, \bar\psi(\gamma_0 + \gamma_1 ) \psi\,\,\rightarrow\,\,J_+^B\,= \, - \,
{1\over{2\pi}}\,\partial_+ \Phi\,
\nonumber\\
J_-^F&=& {1\over \sqrt{2}} \bar\psi (\gamma_0 -\gamma_1 ) \psi \,\,\rightarrow \,\,J_-^B=\, {1\over{2\pi}} \partial_- \, \Phi
\end{eqnarray}

As already commented, the introduction of the compensating field $g$ in the fermionic action made it possible to extract the anomalous divergence of the
chiral current. In the non-Abelian version, this has been given
in Eq. (\ref{17}), which Abelian limit  in light cone coordinates can be written as

\begin{equation}
\label{42}
<\partial_\mu\, J_R^\mu\, >\,\,=\,\,
<\partial_+\,  J_-^F\, >\,=\,- {1\over{ 2\pi }} < \partial_- A_+ \,-\,\partial_+ A_-  >
\end{equation}

\noindent obtained by imposing the independence of the vacuum functional with respect to
$\Lambda$. The corresponding condition applied to the bosonized action (\ref{39}) gives

\begin{equation}
\label{43}
\partial_+ J^B_- \, =\,- \,{1\over{2\pi }}\, (\, \partial_- A_+
\,-\,\partial_+ A_- )
\end{equation}

\noindent which is consistent with the mapping (\ref{41}). Observe that
$\partial_ -J^B_+ + \partial_+ J^B_- $ vanishes identically, since the vector current is  conserved. This is expected, since the regularization was done by choosing the vector symmetry as a preferential one. As a final comment, it is interesting to observe that Eq. (\ref{43}) is just the equation of motion for the field $ \Phi $ in the usual bosonized Schwinger model, when one uses
the definitions of the bosonic chiral currents. This
gives the interpretation that the equation of motion for the scalar field  in the bosonized version of the Schwinger model is just the bosonized version of the expectation value of the anomalous divergence of the $QED_2 $ axial current.

We will show in the next section that similar features also appear when the corresponding non-Abelian models are  considered.

\section{Non-Abelian extension}
\setcounter{equation}{0}
Consider now the  non-Abelian action (\ref{3}). We can write out the bosonized actions that  correspond to each of the chiral sectors as

\begin{eqnarray}
S_+ &=& {1\over 4\pi}\,\int d^2 x \, tr \,\Big[ \, \partial_+ u^{- 1} \partial_- u
\,-\,2i B_+u^{- 1} \partial_- u
\,+\,a B_+ B_- \,\Big] \,+\,\Gamma [ u ]\nonumber\\
S_- &=& {1\over 4\pi}\,\int d^2x \, tr \,\Big[ \, \partial_+ v^{- 1} \partial_- v
\,-\,2i A_- v^{- 1} \partial_+ v
\,+\,b A_+ A_- \, \Big] \,-\,\Gamma [ v ]\,\,,\nonumber\\
& &
\end{eqnarray}

\noindent where again the functional $\Gamma $ is defined as in eq. (\ref{gamma}) and $u$ and $v$  are  elements of
the $SU(N)$ group.

As  in the Abelian case, we solder the two chiralities by introducing the soldering fields $E_+\,$ , $\,E_-\,$ and defining a new action 

\begin{equation}
\label{action}
S\,=\,S_+ [u]\,+\,S_- [v] \,-\, \int d^2x \,\,tr \Big[ E_- {\cal J}_+ \,+\,E_+ 
{\cal J}_- 
\,+\, {1\over 2\pi} E_+ E_- \Big]
\end{equation}

\noindent where

\begin{eqnarray}
{\cal J}_+  &=& {1\over 2\pi} \Big[ u\partial_+ u^{-1}\,-\,i u B_+ u^{-1} \Big]
\nonumber\\
{\cal J}_-  &=& {1\over 2\pi} \Big[ v \partial_- v^{-1}\,-\,i v A_- v^{-1}
 \Big]\,.
\end{eqnarray}

\noindent Action (\ref{action}) is invariant under the transformations 

\begin{eqnarray}\label{4.4}
\delta A_\mu &=& \delta g \,=\,0\nonumber\\
\delta u &=& w u\nonumber\\
\delta v &=& w v\nonumber\\
\delta E_\pm &=& \partial_\pm w \,- [\,E_\pm\,,\,w\,]\,\,.
\end{eqnarray}

\noindent since (\ref{4.4}) imply that

\begin{equation}  
\delta {\cal J}_{\pm}\,=\, - {1\over 2\pi} \partial_\pm\,w \,+\,[\,w\,,\,{\cal J}_\pm\,]\,.
\end{equation}

\noindent Eliminating $ E_\pm \,=\,- 2\pi {\cal J}_\pm\,$ with the aid of their  equations of motion 
 and introducing $ h\,=\,u^{-1} v \,$, we write (\ref{action}) as

\begin{eqnarray}
S' &=& {1\over 4\pi} \int d^2x \,tr \Big[ \, \partial_+ h^{- 1} \partial_- h
\,-\,2i g^{-1} A_+ g h \partial_- h^{-1}
\,+\, 2 g^{-1} \partial_+ g h \partial_- h^{-1}\nonumber\\
&-&2i A_- h^{-1}\partial_+ h 
\,+\, (a\,+\,b) A_+ A_- \,-\,ia A_+ g \partial_- g^{-1}
\,+\, i a \partial_+ g  g^{-1} A_-
\nonumber\\&+& a \partial_+ g^{- 1} \partial_- g  
\,+\, 2\,g^{-1} A_+ g h A_- h^{-1} \,-2i
g^{-1} \partial_+ g h A_- h^{-1}
\,\Big] \,-\,\Gamma [ h ]\nonumber\\
\end{eqnarray}

\noindent Again as in the Abelian model,  vector symmetry  and unitarity are preserved for $a\,=\,b\,=\,1\,$, which is the choice we are going to assume.

The quantum action will also involve the counterterm $\,M_1\,$ of equation (\ref{16}). If we introduce the  new field  $\,G\,=\,gh\,$ we find 

\begin{eqnarray}
\label{QA}
W &=& S'\,+\,M_1 \, = \, {1\over 4\pi} \,\int d^2x tr \Big[ \partial_+ G^{-1} \partial_-G \,-\,2iA_+ G \partial_- G^{-1}\nonumber\\
&-& 2i A_- G^{-1} \partial_+ G \,-\, 2 A_+ G A_- G^{-1} \,+2 A_+ A_- \,\Big]
\,-\Gamma [G]
\end{eqnarray}

\noindent that has the same form as the bosonized version of $QCD_2$ action. So, we see that the compensating field $\,g\,$ shows up in the bosonized non-Abelian case also as a collective field. This of course was only possible due to the perfect matching between $S'$ and $M_1$.

In order to find the correct mapping among the fermionic and  bosonic currents  we recall that the non-Abelian version of the definition (\ref{Acurr}) is

\begin{equation}
J_\pm^a  \,=\,{ \delta W \over \delta A_\mp^a }\vert_{_{ g = {\bf 1}}}\,\,\,\,. 
\end{equation}

\noindent That corresponds in the fermionic formulation to the trivial non-Abelian version of eq. (\ref{41}) and in the bosonized formulation, using the quantum action $W$ of equation 
(\ref{QA}) to

\begin{eqnarray}
J_+ &\equiv& J_+^a T^a\,=\, {1\over 2\pi} \Big( - i h^{-1} \partial_+ h \,-\, h^{-1} A_+ h \,+\,A_+ \Big)\nonumber\\
J_{-} &\equiv& J_-^a T^a\,=\, {1\over 2\pi} \Big( - i h \partial_- h^{-1}  \,-\,h A_- h^{-1}   \,+\,A_-\Big)\,\,\,.
\end{eqnarray}

\noindent Note that, differently from the fermionic case, the current involves the gauge field itself, which is a consequence of the self-interaction presented by the non-Abelian model.

Now, in order to 
extract the result corresponding to (\ref{17}), we choose

\begin{equation}
G = ({\bf1}\,+\,i\beta) h \nonumber\\
\end{equation}

\noindent  with $\,\beta\,=\,\beta^a T^a\,$ small. At this gauge we find

\begin{equation}\label{61}
{\delta \over \delta \beta^a} \,W\vert_{_{\beta = 0}}\,=\,
(\,D_+ J_-\,)^a \,-\,{1\over 2\pi} (\partial_+ A_-^a \,-\, \partial_- A_+^a \,)
\end{equation}

Therefore, independence of the theory with respect to $\beta$ gives

\begin{equation}
\label{62}
(\,D_+ J_-\,)\,=\,-\, {1\over 2\pi} ( \partial_- A_+\, \,-\, \partial_+ A_- )
\end{equation}

\noindent which is the expected result for the anomalous divergency of the chiral current in  $QCD_2$, here reproduced in the bosonized formulation.

\section{Conclusions}
We have shown in this work how to find a bosonization scheme compatible with the introduction of compensating fields in $QCD_2$. The mapping between the matter currents in the fermionic and bosonic formulations was defined in an unambiguous way, by 
looking at the coupling to the gauge field. We have seen that the bosonic currents involve a non trivial gauge field dependent contribution that is not present in the fermionic description. 

Only by including in the bosonic formulation the counterterm that comes from the master equation at one loop order in the fermionic description we have the same set of symmetries in both cases. 
It is  crucial at this point the fact that this counterterm does not depends on  fermionic variables and so does  not need to be bosonized.
Also only by the inclusion of those quantum corrections to the action, the anomalous divergence of the bosonized chiral current
could be properly derived from the equations of motion of the compensating fields.

\vskip 1cm
\noindent {\bf Acknowledgment:} We are indebted to R.L.P.G. Amaral for an important discussion. This work is partially supported  by
 CNPq, FINEP and FUJB (Brazilian Research Agencies).

\end{document}